\documentclass[12pt]{article}
\usepackage{epsf}
\usepackage{amsmath,amssymb}
\usepackage{latexsym}
\usepackage{graphicx,color}
\usepackage{wrapfig}
\usepackage{latexsym}
\usepackage{graphics}
\usepackage{cite}

\usepackage{color}
\usepackage{delarray,color}
\usepackage{fancybox}  
\newcommand {\beq}{\begin{eqnarray}}
\newcommand {\eeq}{\end{eqnarray}}

\usepackage{tabularx}

\catcode`\@=11
\@addtoreset{equation}{section}

\catcode`@=12
\relax

\newcommand{\ep}{\epsilon}
\newcommand{\ti}{\times}
\newcommand{\tr}{\mathrm{Tr}}

\newcommand{\llangle}{\langle \!\langle}
\newcommand{\rrangle}{\rangle \!\rangle}

\begin{document}
\baselineskip 0.7cm

\begin{titlepage}

\setcounter{page}{0}

\renewcommand{\thefootnote}{\fnsymbol{footnote}}

\begin{flushright}
YITP-08-36\\
UT-08-15
\end{flushright}

\vskip 1.35cm

\begin{center}
{\Large \bf 
A New $\mathcal{N}=4$ Membrane Action via Orbifold
}

\vskip 1.2cm 

{\normalsize
Hiroyuki Fuji$^1$\footnote{hfuji(at)yukawa.kyoto-u.ac.jp}, Seiji Terashima$^1$\footnote{terasima(at)yukawa.kyoto-u.ac.jp} and Masahito~Yamazaki$^{1,2}$\footnote{yamazaki(at)hep-th.phys.s.u-tokyo.ac.jp}
}

\vskip 0.8cm

{ \it
$^1$Yukawa Institute for Theoretical Physics, Kyoto University, Kyoto 606-8502, Japan
\\[2mm]
$^2$Department of Physics, University of Tokyo, Hongo 7-3-1, Tokyo, 113-0033, Japan
}

\end{center}

\vspace{12mm}

\centerline{{\bf Abstract}}
We propose a new Lagrangian describing $\mathcal{N}=4$ superconformal field theory in three dimensions. This theory is believed to describe interacting field theory on the worldvolume of a M2-brane on an orbifold, and is obtained as a $\mathbb{Z}_2$-quotient of the theory proposed by Bagger and Lambert. 
Despite unusual Chan-Paton structures, we can take $\mathbb{Z}_2$-orbifold by using $SU(2)\times SU(2)$ bifundamental representations. 
We also analyze the moduli space of this theory and found three 
branches. 
With an assumption of a broken $U(1)$ symmetry,
the moduli space is consistent with that of the D2-brane 
in the strong coupling limit of Type IIA string theory 
if the gauge group is $O(4)$.
Our action has manifest $\mathbb{Z}_2$-symmetry exchanging two $\mathbb{R}^4/\mathbb{Z}_2$'s in M-theory, and this suggests a new non-perturbative duality between a O2$^-$-brane on orbifold $\mathbb{R}^4/\mathbb{Z}_2$ and a O2$^-$-brane with D6-branes. 

\end{titlepage}
\newpage

\section{Introduction}
In \cite{BL2}, motivated by early attempts \cite{Schwarz,BH}, Bagger and Lambert proposed a new Lagrangian description of three-dimensional maximally supersymmetric ($\mathcal{N}=8$)  conformal field theory with manifest $SO(8)$-symmetry (see also \cite{BL1,G,BL3}). The theory is believed to be realized on the worldvolume of multiple M2-branes in M-theory, and many aspects of the theory has been explored recently \cite{vR,M2toD2,BLS,Berman,Morozov,LT,Mfold,Gomis,Bergshoeff,Hosomichi,Gran,HHM,Papadopoulos1,GG,Papadopoulos2,HM,Gomis2,Benvenuti,HIM,Morozov2,Iso}.

Despite their success, we have so far only a single example of interacting field theories on the worldvolume of membranes, the so-called $\mathcal{A}_4$-theory, which is interpreted as the worldvolume theory of two M2-branes in M-theory on $\mathbb{R}^8/\mathbb{Z}_2$ \cite{LT,Mfold}. The original construction in \cite{BL2} was based on new algebraic structures called Lie 3-algebras (and non-associative algebra), and there was hope for some time that there might exist many other Lie 3-algebras. However, it later was conjectured \cite{HHM} and later proven \cite{Papadopoulos1,GG} that only the $\mathcal{A}_4$-theory is allowed in the framework of \cite{BL1} under the condition of the positivity of the metric.\footnote{By abandoning positivity we can construct more examples of theories \cite{Gran,HHM,Gomis2,Benvenuti,HIM,Iso}.} Thus there is a pressing need to have more examples of Lagrangians describing theories on membranes. 

In this paper we propose a new Lagrangian describing three-dimensional $\mathcal{N}=4$ superconformal gauge theory. 
Our theory is obtained as a $\mathbb{Z}_2$-quotient of Bagger-Lambert theory. This is non-trivial because the structure of Chan-Paton factor is unusual in Bagger-Lambert theory. Our study shows that $SU(2)\times SU(2)$ bifundamental representation \cite{vR}, rather than the original $SO(4)$ notation \cite{BL2}, is essential for our purposes. 
Orbifolding also serves as a consistency check of
the proposal that Bagger-Lambert theory describes theories on multiple
M2-branes.
For $\mathbb{Z}_2$-orbifolded Bagger-Lambert theory, we find three 
branches of the moduli space. For Coulomb branches, we assumed the
breaking of the $U(1)$ symmetry to its discrete subgroup
$\mathbb{Z}_m$.\footnote{Within the framework of the Bagger-Lambert
theory, we cannot justify this assumption explicitly. In terms of ABJM theory \cite{ABJM}, 
such breaking is naturally realized and orbifolded moduli space is 
studied \cite{TY}.} 
The consistency with D2-branes picture \cite{M2toD2}
requires that such orbifolds should exist in the strong 
coupling limit and it should describe 
M-theory on $\mathbb{R}^8/(\mathbb{Z}_2 \times \mathbb{Z}_2)$.\footnote{This can also be written as $(\mathbb{R}^4/\mathbb{Z}_2) \times (\mathbb{R}^4/\mathbb{Z}_2)$, and thus we have manifest $\mathbb{Z}_2$-symmetry exchanging two $\mathbb{R}^4/\mathbb{Z}_2$'s. In Type IIA language, this exchanges orientifold and $\mathbb{Z}_2$ orbifold, which is highly non-trivial. We will comment on the significance of this fact in the discussions.}
Actually in the case of $m=4$, the moduli space for the 
D2-brane with O2$^-$-plane on the orbifold is consistent with 
that of $\mathbb{Z}_2$-orbifolded Bagger-Lambert theory.

Another motivation comes from the recent work of \cite{GW2}. Although our theory differs from that of \cite{GW2}, it also discusses three-dimensional Chern-Simons theories with $\mathcal{N}=4$ supersymmetry, which is similar to our theory in many respects.

The organization of this article is as follows. We begin in section \ref{sec:BLreview} with a brief summary of Bagger-Lambert theory in $SU(2)\times SU(2)$ bifundamental representation. Next we discuss in section \ref{sec:orbifold} the $\mathbb{Z}_2$-quotient of Bagger-Lambert theory. Then in section \ref{sec:moduli} we discuss the moduli space of theory. Section \ref{sec:conclusion} is devoted to conclusions and discussions. In appendix \ref{sec:gamma} we summarize our notations of $\Gamma$-matrices.

\section{Bagger-Lambert theory in bifundamental representation} \label{sec:BLreview}

In this section, in order to set up notations used in this paper, we briefly review the Bagger-Lambert theory \cite{BL2} using the $SU(2)\times SU(2)$ bifundamental notation of \cite{vR}. Although the original paper \cite{BL2} uses the $SO(4)$ notation, $SU(2)\times SU(2)$ notation is essential for our purposes.

The matter contents of the theory consists of eight scalar fields $X^{I}$ ($I=1,\ldots ,8$), 11-dimensional Majorana fermion $\Psi$, and two gauge fields $A_{\mu}$ and $\hat{A}_{\mu}$.
In bifundamental representation, the scalar fields $X^{I}$ and fermionic
fields $\Psi$ are represented by a 2 $\times$ 2 matrix
\begin{eqnarray}
X^{I}=\frac{1}{2}\left(
\begin{array}{cc}
x_4^I+ix_3^I & x_2^I+ix_1^I\\
-x_2^I+ix_1^I & x_4^I-ix_3^I 
\end{array}
\right),
\quad
\Psi=\frac{1}{2}\left(
\begin{array}{cc}
\psi_4+i\psi_3 & \psi_2+i\psi_1\\
-\psi_2+i\psi_1 & \psi_4-i\psi_3 
\end{array}
\right),
\label{matrix form}
\end{eqnarray}
and similarly for gauge fields
\begin{align}
i A_{\mu}=\left(
\begin{array}{cc}
ia^{\mu} & a_2^{\mu}+ia_1^{\mu}\\
-a_2^{\mu}+ia_1^{\mu} & -ia^{\mu} 
\end{array}
\right), ~~
i\hat{A}_{\mu}=\left(
\begin{array}{cc}
i\hat{a}^{\mu} & \hat{a}_2^{\mu}+i\hat{a}_1^{\mu}\\
-\hat{a}_2^{\mu}+i\hat{a}_1^{\mu} & -i\hat{a}^{\mu} 
\end{array}
\right).
\end{align}
Note that gauge fields are represented by traceless matrices, and their diagonal components are written as $a^{\mu}$ and $\hat{a}^{\mu}$, rather than $a_3^{\mu}$ and $\hat{a}_3^{\mu}$, respectively.
%
The reality conditions for $X_I$'s are given by
\begin{align}
X_{\alpha \dot{\beta}}=\epsilon_{\alpha \beta} \ep_{\dot{\beta} \dot{\alpha}} \left( X^{\dagger} \right)^{\dot{\alpha} \beta},
\end{align}
and we also have the chirality condition for $\Psi$:
\begin{align}
\Gamma^{012}\Psi=-\Psi.
\end{align}

In this notation, the Lagrangian of the Bagger-Lambert theory is given by
\begin{align}
{\cal L} &=  \tr\left( -(D^\mu X^I)^\dagger D_\mu X^I + i \bar{\Psi}^{\dagger} \Gamma^{\mu} D_\mu \Psi \right)\cr
& + \tr\left(-{2 \over 3} if\, \bar{\Psi}^\dagger \Gamma_{IJ} (X^I X^{J \dagger} \Psi + X^J \Psi^\dagger X^I + \Psi X^{I \dagger} X^J)  - {8 \over 3}f^2 X^{[I} X^{J \dagger} X^{K]}X^{K \dagger} X^J X^{I \dagger}\right) \cr
& + {1 \over 2f} \epsilon^{\mu \nu \lambda} \tr\left(A_\mu \partial_\nu A_\lambda + {2 \over 3}i A_\mu A_\nu A_\lambda\right)  - {1 \over 2f} \epsilon^{\mu \nu \lambda} \tr\left(\hat{A}_\mu \partial_\nu \hat{A}_\lambda + {2 \over 3}i \hat{A}_\mu \hat{A}_\nu \hat{A}_\lambda\right),
\end{align}
where the covariant derivative is defined by
\beq
D_{\mu} X^I=\partial_{\mu} X^I+i A_{\mu} X^I-i X^I \hat{A}_{\mu}.
\eeq

The supersymmetry transformations, under which the action is invariant, are given by
\begin{eqnarray}
&&\delta X^I=i\bar{\epsilon}\Gamma^I\Psi,
\\
&&\delta \Psi=D_{\mu}X^I\Gamma^{\mu}\Gamma^I\epsilon
+\frac{2}{3}fX^IX^{J\;\dagger}X^K\Gamma^{IJK}\epsilon,
\\
&&\delta A_{\mu}=f\bar{\epsilon}\Gamma_{\mu}\Gamma_I (X^I\Psi^{\dagger}-\Psi X^{I\;\dagger}),
\\
&&\delta \hat{A}_{\mu}=f\bar{\epsilon}\Gamma_{\mu}\Gamma_I (\Psi^{\dagger}X^I- X^{I\;\dagger}\Psi),
\end{eqnarray}
where the spinor $\epsilon$ has the opposite chirality from $\Psi$:
\beq
\Gamma^{012}\ep=\ep.
\eeq

Finally, in order to make the action invariant under large coordinate transformations, the parameter $f$ should take the form
\beq
f=\frac{2\pi}{k},
\eeq
where the level $k$ is a positive integer.

\section{$\mathbb{Z}_2$-action and its invariant sector}\label{sec:orbifold}
In this section we shall consider the $\mathbb{Z}_2$-quotient 
of the Bagger-Lambert theory.
We consider a discrete group $\mathbb{Z}_2$ acting on $\mathbb{R}^4$ in the $\mathbb{R}^8$ spatial directions transverse to M2-branes. We therefore decompose the eight scalar fields $X^I$
($I=1,\cdots,8$) into $Z^i$ ($i=1,\cdots,4$) and $Y^s$ ($s=5,\cdots,8$).
For each field our $\mathbb{Z}_2$ acts as follows:
\begin{equation}
 Z^i \to -\gamma Z^i \gamma, ~~ Y^s \to \gamma Y^s \gamma, ~~
\Psi \to \Gamma^{1234} \gamma \Psi \gamma,~~
A_{\mu} \to \gamma A_{\mu}\gamma, \quad 
   \hat{A}_{\mu} \to \gamma \hat{A}_{\mu}\gamma,  
\label{eq:Z2}
\end{equation}
where $\gamma$ is the regular representation of $\mathbb{Z}_2$ given by
\begin{eqnarray}
\gamma=\left(
\begin{array}{cc}
1 & 0\\
0 & -1
\end{array}
\right).
\end{eqnarray}
This matrix $\gamma$ is chosen so that it is consistent with the usual discussions of orbifolds \cite{DM} after the reduction to (the strong coupling limit of) D2-branes \cite{M2toD2}.
For the fermionic field $\Psi$ the quotient action is realized as 
the $\Gamma^{1234}:=\Gamma^1\Gamma^2\Gamma^3\Gamma^4$ action. This corresponds to $\mathbb{Z}_2$-action on $\mathbb{R}^4$ in $\mathbb{R}^8$, or $\pi$ rotations in both $12$ and $34$ directions. The details are explained in the appendix.

For $Z^i$, $Y^s$ and $\Psi$, the $\mathbb{Z}_2$-quotient acts
simply as multiplications by $\pm 1$ on their diagonal (D) and off-diagonal (A) parts: 
\begin{eqnarray}
&& Z^i=Z_D^i+Z_A^i,\quad Y^s=Y_D^s+Y^s_A,\label{eq:DAdecomp}\\
&&
Z_D^i\to 
-Z_D^i,\quad 
Z_A^i\to 
Z_A^i,\quad 
Y_D^s\to 
Y_D^s,\quad 
Y_A^s \to 
-Y_A^s,\label{eq:DAtransf}
\end{eqnarray}


The fermionic fields should be further decomposed into
$\Gamma^{1234}$ eigenstates
\begin{eqnarray}
&&\Psi=\Psi_{D}+\Psi_{A}=\Psi_{D+}+\Psi_{D-}+\Psi_{A+}+\Psi_{A-},
\\
&& \Psi_{D\pm}=P_{\pm} \Psi_D,\quad 
   \Psi_{A\pm}=P_{\pm} \Psi_A,\quad
\Psi_{D\pm}\to \pm \Psi_{D\pm},\quad 
\Psi_{A\pm}\to \mp \Psi_{A\pm},
\end{eqnarray}
where
\begin{equation}
P_{\pm}:=\frac{1}{2}(1\pm\Gamma^{1234}),
\end{equation}
are the projectors onto $\Gamma^{1234}=\pm 1$.

\subsection{Orbifold by $\mathbb{Z}_2$}
Now we would like to prove that the $\mathbb{Z}_2$-truncation as given by \eqref{eq:Z2} gives a consistent theory with $\mathcal{N}=4$ supersymmetry. To begin with, we discuss conditions under which $\mathcal{N}=4$ supersymmetry is preserved after the $\mathbb{Z}_2$-truncation. 

We first decompose the fields into
the two types: the $\mathbb{Z}_2$-invariant fields 
\begin{eqnarray}
{\cal I}=\{ Z_A, Y_D, \Psi_{D+},\Psi_{A-},A_D,\hat{A}_D  \},
\label{calI}
\end{eqnarray}
and the other fields 
\begin{eqnarray}
{\cal N}=\{ Z_D, Y_A, \Psi_{D-},\Psi_{A+},A_A,\hat{A}_A  \},
\end{eqnarray}
which will be projected out.
The action of the orbifolded theory will be defined by
\begin{eqnarray}
\tilde{S}({\cal I})=S({\cal I},{\cal N}) |_{{\cal N}=0},
\end{eqnarray}
from the original action $S({\cal I},{\cal N})$.
Then the symmetry $\delta$ of the original theory 
will become also a symmetry of the orbifolded theory 
if the following condition is satisfied:
\begin{eqnarray}
\delta {\cal N} |_{{\cal N}=0}=0.
\label{cond1}
\end{eqnarray}
In such a case the symmetry of the orbifolded theory is generated by  
\begin{eqnarray}
\tilde{\delta} {\cal I} = \delta {\cal I} |_{{\cal N}=0}.
\end{eqnarray}
Indeed, from $\delta S=0$ we can easily show that 
\begin{eqnarray}
\tilde{\delta} \tilde{S} = 0,
\label{dS}
\end{eqnarray}
by expansion with respect to ${\cal N}$.

\subsection{Compatibility of $\mathbb{Z}_2$-orbifold with $\mathcal{N}=4$ supersymmetry}
Let us now examine condition \eqref{cond1} to ensure that we have remaining $\mathcal{N}=4$ supersymmetry.
From the definition of $Z_D$ and $\gamma$, 
$Z_D := (Z+\gamma Z \gamma)/2$ and we find
\begin{eqnarray}
\delta Z_D^i =\frac{1}{2} \left(\delta Z^i+\gamma (\delta Z^i) \gamma\right)
=i\bar{\epsilon}\Gamma^i \Psi_{D}.
\end{eqnarray}
Thus, 
\begin{eqnarray}
\delta Z_D^i |_{{\cal N}=0} 
=i\bar{\epsilon}\Gamma^i \Psi_{D+}
=i\bar{\epsilon}\Gamma^i  P_+ \Psi_{D+}
=i\bar{\epsilon} P_- \Gamma^i \Psi_{D+},
\end{eqnarray}
and the (\ref{cond1}) implies that 
the surviving supersymmetry should satisfy a chirality condition
\begin{eqnarray}
P_{-}\epsilon=\frac{1}{2}(1-\Gamma^{1234})\epsilon=0.
\label{chi}
\end{eqnarray}
We also find 
\begin{eqnarray}
\delta Y_A^s |_{{\cal N}=0} 
=i\bar{\epsilon}\Gamma^s \Psi_{A-},
\end{eqnarray}
which will vanish with (\ref{chi}).

The supersymmetry transformations for $\Psi_{D-}$ and $\Psi_{A+}$ are
\begin{eqnarray}
\delta\Psi_{D-}|_{{\cal N}=0}
&=&\delta\Bigl[P_-\Psi_{D-}\Bigr]\Big|_{{\cal N}=0}
\nonumber \\
&=&\Bigl[(\partial_{\mu}Y_D^s
+iA_{\mu D}Y_D^s-iY_D^s\hat{A}_{\mu D})\Gamma^{\mu}\Gamma^s
+\frac{2}{3}fY_D^sY_D^{t\;\dagger}Y_D^u\Gamma^{stu} 
\nonumber \\
&&
 +\frac{2}{3}f(Y_D^sZ_A^{i\;\dagger}Z_A^j
+Z_A^jY_D^{s\;\dagger}Z_A^i
+Z_A^iZ_A^{j\;\dagger}Y_D^s)
\Gamma^{ijs}\Bigr ]P_-\epsilon,
\\
\delta\Psi_{A+}|_{{\cal N}=0}
&=&\delta\Bigl[P_+\Psi_{A+}\Bigr]\Big|_{{\cal N}=0}
\nonumber \\
&=&\Bigl[(\partial_{\mu}Z_A^i
+iA_{\mu D}Z_A^i-iZ_A^i\hat{A}_{\mu D})\Gamma^{\mu}\Gamma^i
+\frac{2}{3}fZ_A^iZ_A^{j\;\dagger}Z_A^k\Gamma^{ijk}
\nonumber \\
&&
+\frac{2}{3}f(Z_A^iY_D^{s\;\dagger}Y_D^t
+Y_D^tZ_A^{i\;\dagger}Y_D^s
+Z_A^sZ_A^{t\;\dagger}Z_A^i)
\Gamma^{sti}\Bigr]P_-\epsilon.
\end{eqnarray}
Thus we also find $\delta{\cal N}|_{{\cal N}=0}=0$ 
for the fermionic fields if (\ref{chi}) is satisfied. It is also easy to check 
the compatibility condition for gauge fields. In this way we have proven that $\mathcal{N}=4$ supersymmetry is preserved after the truncation.

\subsection{The Lagrangian and its remaining $\mathcal{N}=4$ supersymmetry}

The surviving supersymmetry transformations are summarized as follows.
\begin{eqnarray}
\tilde{\delta}Z_A^i
&=& i\bar{\epsilon}\Gamma^i\Psi_{A-},\\
\tilde{\delta}Y_D^s
&=& i\bar{\epsilon}\Gamma^s\Psi_{D+}, \\
\tilde{\delta} \Psi_{D+}
&=& (\partial_{\mu}Y_D^s
+iA_{\mu D}Y_D^s-iY_D^s\hat{A}_{\mu D})\Gamma^{\mu}\Gamma^s\epsilon
\nonumber \\
&&
+\frac{2}{3}f(Y_D^sZ_A^{i\dagger}Z_A^j
+Z_A^j{Y}_D^{s\dagger}Z_A^i
+Z_A^iZ_A^{j\dagger}Y_D^s)
\Gamma^{ijs}\epsilon,
\\
\tilde{\delta} \Psi_{A-}
&=&(\partial_{\mu}Z_A^i
+iA_{\mu D}Z_A^i-iZ_A^i\hat{A}_{\mu D})\Gamma^{\mu}\Gamma^i\epsilon
\nonumber \\
&&
+\frac{2}{3}f(Z_A^i{Y}_D^{s\dagger}Y_D^t
+Y_D^tZ_A^{i\dagger}Y_D^s
+Z_A^sZ_A^{t\dagger}Z_A^i)
\Gamma^{sti}\epsilon, \\
\tilde{\delta}A_{\mu D}
&=&
f\bar{\epsilon}\Gamma_{\mu}\Gamma_i
(Z_A^i\Psi_{A-}^{\dagger}-\Psi_{A-}Z_A^{i\dagger})
+f\bar{\epsilon}\Gamma_{\mu}\Gamma_s
(Y_D^s\Psi_{D+}^{\dagger}-\Psi_{D+}Y_D^{s\dagger}),
 \\
\tilde{\delta}\hat{A}_{\mu D}
&=&
f\bar{\epsilon}\Gamma_{\mu}\Gamma_i
(\Psi_{A-}^{\dagger}Z_A^i-Z_A^{i\dagger}\Psi_{A-})
+f\bar{\epsilon}\Gamma_{\mu}\Gamma_s
(\Psi_{D+}^{\dagger}Y_D^{s}-Y_D^{s\dagger}\Psi_{D+}).
\end{eqnarray}
In components, the supersymmetry transformations are
\begin{eqnarray}
&& \tilde{\delta}z_1^i=i\bar{\epsilon}\Gamma^i\psi_1,\quad
\tilde{\delta}z_2^i=i\bar{\epsilon}\Gamma^i\psi_2,
\quad
 \tilde{\delta}y_3^s=i\bar{\epsilon}\Gamma^s\psi_3,\quad
\tilde{\delta}y_4^s=i\bar{\epsilon}\Gamma^s\psi_4,
\\
&&
\tilde{\delta}\psi_1
=\left[\partial_{\mu}z_1^i+(a_{\mu}+\hat{a}_{\mu})z_2^i\right]
\Gamma^{\mu}\Gamma^s\epsilon
+\frac{1}{2}fz_2^i(y_3^sy_4^t-y_3^ty_4^s)\Gamma^{sti}\epsilon,
\\
&&
\tilde{\delta}\psi_2
=\left[\partial_{\mu}z_2^i-(a_{\mu}+\hat{a}_{\mu})z_1^i\right]
\Gamma^{\mu}\Gamma^s\epsilon
-\frac{1}{2}fz_1^i(y_3^sy_4^t-y_3^ty_4^s)\Gamma^{sti}\epsilon,
\\
&& 
\tilde{\delta}\psi_3
=\left[\partial_{\mu}y_3^s+(a_{\mu}-\hat{a}_{\mu})y_4^s\right]
\Gamma^{\mu}\Gamma^s\epsilon
+\frac{1}{2}fy_4^s(z_1^iz_2^j-z_2^iz_1^j)\Gamma^{ijs}\epsilon.
\\
&&
\tilde{\delta}\psi_4
=\left[\partial_{\mu}y_4^s-(a_{\mu}-\hat{a}_{\mu})y_3^s\right]
\Gamma^{\mu}\Gamma^s\epsilon
-\frac{1}{2}fy_3^s(z_1^iz_2^j-z_2^iz_1^j)\Gamma^{ijs}\epsilon,
\\
&&\tilde{\delta}a_{\mu}=
\frac{i}{2}f\bar{\epsilon}\Gamma_{\mu}\Gamma_i(z_1^i\psi_2-z_2^i\psi_1)
+\frac{i}{2}f\bar{\epsilon}\Gamma_{\mu}\Gamma_s(y_3^s\psi_4-y_4^s\psi_3),
\\
&&\tilde{\delta}\hat{a}_{\mu}=
-\frac{i}{2}f\bar{\epsilon}\Gamma_{\mu}\Gamma_i(z_1^i\psi_2-z_2^i\psi_1)
+\frac{i}{2}f\bar{\epsilon}\Gamma_{\mu}\Gamma_s(y_3^s\psi_4-y_4^s\psi_3),
\end{eqnarray}

The Lagrangian for $\mathbb{Z}_2$-orbifolded theory is
\footnote{We multiplied $1/2$ 
factor to the Lagrangian
in order to reproduce correct membrane tension in the Coulomb branch
\cite{TY}.}
\begin{eqnarray}
{\cal L}&=&
\frac{1}{2}
{\rm Tr}\left[-({\cal D}^{\mu}_DY_D^s)^{\dagger}({\cal D}_{\mu D}Y_D^s)
-({\cal D}^{\mu}_DZ_A^i)^{\dagger}({\cal D}_{\mu D}Z_A^i)\right] 
\nonumber \\
&&+\frac{i}{2}
{\rm Tr}
\left[\bar{\Psi}^{\dagger}_{D+}\Gamma^{\mu} {\cal D}_{\mu D}\Psi_{D+}
+\bar{\Psi}^{\dagger}_{A-}\Gamma^{\mu}{\cal D}_{\mu D}\Psi_{A-}
\right]
\nonumber \\
&&
-if{\rm Tr}\Bigl[
\bar{\Psi}^{\dagger}_{D+}\Gamma_{ij}
\llangle Z_A^i,Z_A^{j\dagger},\Psi_{D+} \rrangle
\Bigr]
-if{\rm Tr}\Bigl[
\bar{\Psi}^{\dagger}_{A-}\Gamma_{st}
\llangle Y_D^s,Y_D^{t\dagger},\Psi_{A-} \rrangle
\Bigr]
\nonumber \\
&&
-i
f{\rm Tr}\Bigl[
\bar{\Psi}^{\dagger}_{D+}\Gamma_{si}
\llangle Y_D^s,Z_A^{i\dagger},\Psi_{A-} \rrangle
\Bigr]
-i
f{\rm Tr}\Bigl[
\bar{\Psi}^{\dagger}_{A-}\Gamma_{si}
\llangle Y_D^s,Z_A^{i\dagger},\Psi_{D+} \rrangle\Bigr]
\nonumber \\
&&-
\frac{1}{2}
V(Z_A,Y_D)
+
\frac{1}{4f}
\epsilon^{\mu\nu\rho}
{\rm Tr}\left[A_{\mu D}\partial_{\nu}A_{\rho D}
-\hat{A}_{\mu D}\partial_{\nu}\hat{A}_{\rho D}\right],
\end{eqnarray}
where the covariant derivative $\mathcal{D}_D$ is defined by (when acting on $Y_D^s$, for example)
\begin{align}
\mathcal{D}_{\mu D}Y_D^s=\partial_{\mu}Y_D^s+i A_{\mu D}Y_D^s -i Y_D^s \hat{A}_{\mu D},
\end{align}
and the potential $V(Z_A,Y_D)$ is given by
\begin{eqnarray}
V(Z_A,Y_D)&=&
\frac{8}{3}f^2{\rm Tr}\Bigl[\llangle Y_D^{s},Z_A^{i\dagger},Z_A^{j} \rrangle Z_A^{j\dagger}Z_A^{i}Y_D^{s\dagger}
+\llangle Z_A^{j},Y_D^{s\dagger},Z_A^{i} \rrangle Z_A^{i\dagger}Y_D^{s}Z_A^{j\dagger} \nonumber\\
&&\hspace{1cm}+\llangle Z_A^{i},Z_A^{j\dagger},Y_D^{s} \rrangle Y_D^{s\dagger}Z_A^{j}Z_A^{i\dagger}
+\llangle Z_A^{i},Y_D^{s\dagger},Y_D^{t} \rrangle Y_D^{t\dagger}Y_D^{s}Z_A^{i\dagger}
\nonumber \\
&&\hspace{1cm}+\llangle Y_D^{t},Z_A^{i\dagger},Y_D^{s}\rrangle Y_D^{s\dagger}Z_A^{i}Y_D^{t\dagger}
+\llangle Y_D^{s},Y_D^{t\dagger},Z_A^{i}\rrangle Z_A^{i\dagger}Y_D^{t}Y_D^{s\dagger} \Bigr]
\nonumber \\
&=&\frac{1}{4}f^2\left[
((y_3^{s})^2+(y_4^{s})^2)(z_1^jz_2^i-z_1^iz_2^j)^2
+((z_1^{i})^2+(z_2^{i})^2)(y_3^ty_4^s-y_3^sy_4^t)^2\right].
\nonumber \\
\label{potential}
\end{eqnarray}
In these equations $\llangle ~~\rrangle$ stands for summation over signed permutations with position of dagger fixed. For example,

\begin{align}
\begin{split}
\llangle Z_A^i,Z_A^{j\dagger},\Psi_{D+} \rrangle:=
\frac{1}{6}\Bigl(&Z_A^iZ_A^{j\dagger}\Psi_{D+}
+Z_A^j\Psi^{\dagger}_{D+}Z_A^{i}
+\Psi_{D+}Z_A^{i\dagger}Z_A^{j}\\
&-Z_A^jZ_A^{i\dagger}\Psi_{D+}
-Z_A^i\Psi^{\dagger}_{D+}Z_A^{j}
-\Psi_{D+}Z_A^{j\dagger}Z_A^{i}
\Bigr), \\
\end{split}
\end{align}
and
\begin{align}
\begin{split}
\llangle Y_D^{s},Z_A^{i\dagger},Z_A^{j} \rrangle:=\frac{1}{6}\Bigl(&
Y_D^{s}Z_A^{i\dagger}Z_A^{j}
+Z_A^{j}Y_D^{s\dagger}Z_A^{i}
+Z_A^{i}Z_A^{j\dagger}Y_D^{s} 
\\
&-Z_A^{i}Y_D^{s\dagger}Z_A^{j}
-Y_D^{s\dagger}Z_A^{j\dagger}Z_A^{i}
-Z_A^{j}Z_A^{i\dagger}Y_D^{s} \Bigr).
\end{split}
\end{align}

In terms of components, the Lagrangian is explicitly written down as
follows.
\begin{eqnarray}
{\cal L}&=&-
\frac{1}{4}
\left|[\partial_{\mu}+i(a_{\mu}-\hat{a}_{\mu})](y_4^s+iy_3^s))\right|^2
-\frac{1}{4}
\left|[\partial_{\mu}+i(a_{\mu}+\hat{a}_{\mu})](z_2^i+iz_1^i))\right|^2
\nonumber \\
&&+
\frac{i}{4}\Bigl[
\bar{\psi_1}\partial \hspace{-.60em}/\psi_1
+\bar{\psi_2}\partial \hspace{-.60em}/\psi_2
+\bar{\psi_3}\partial \hspace{-.60em}/\psi_3
+\bar{\psi_4}\partial \hspace{-.60em}/\psi_4
\nonumber \\
&&
+(a_{\mu}+\hat{a}_{\mu})\bar{\psi}_1\Gamma^{\mu}\psi_2
-(a_{\mu}+\hat{a}_{\mu})\bar{\psi}_2\Gamma^{\mu}\psi_1
+(a_{\mu}-\hat{a}_{\mu})\bar{\psi}_3\Gamma^{\mu}\psi_4
-(a_{\mu}-\hat{a}_{\mu})\bar{\psi}_4\Gamma^{\mu}\psi_3\Bigr]
\nonumber \\
&&
+\frac{k^{\prime}}{2\pi}
\epsilon^{\mu\nu\rho}
(a_{\mu}\partial_{\nu}a_{\rho}-\hat{a}_{\mu}\partial_{\nu}\hat{a}_{\rho})
\nonumber \\
&&-\frac{i}{8}
f(z_1^iz_2^j-z_2^iz_1^j)
(\bar{\psi}_3\Gamma_{ij}\psi_4-\bar{\psi}_4\Gamma_{ij}\psi_3)
-\frac{i}{8}
f(y_3^sy_4^t-y_4^sy_3^t)
(\bar{\psi}_1\Gamma_{st}\psi_2-\bar{\psi}_2\Gamma_{st}\psi_1)
\nonumber \\
&&+
\frac{i}{8}
f(y_4^s\bar{\psi}_3-y_3^s\bar{\psi}_4)\Gamma_{si}
(z_2^i\psi_1-z_1^i\psi_2)
+
\frac{i}{8}
f(z_2^i\bar{\psi}_1-z_1^i\bar{\psi}_2)\Gamma_{is}
(y_4^s\psi_3-y_3^s\psi_4)
\nonumber \\
&&
-
\frac{1}{8}
f^2[
((y_3^{s})^2+(y_4^{s})^2)(z_1^jz_2^i-z_1^iz_2^j)^2
+((z_1^{i})^2+(z_2^{i})^2)(y_3^ty_4^s-y_3^sy_4^t)^2].
\label{comp action}
\end{eqnarray}
The Chern-Simons gauge coupling $k^{\prime}$ of the $\mathbb{Z}_2$
orbifolded theory is related with that of the original action as\footnote{The $\mathbb{Z}_2$-orbifolding can only be performed in the case
of even $k$ \cite{TY}.}
\begin{eqnarray}
k^{\prime}=k/2.
\label{orbicoupling}
\end{eqnarray}

\subsection{Discrete symmetries of the Lagrangian} 
By $\mathbb{Z}_2$-orbifolding, the gauge group of our theory is naively broken down to $U(1)\times U(1)$ generated by $a_{\mu}$ and $\hat{a}_{\mu}$. However, we have one discrete gauge symmetry $\mathbb{Z}_2$, which is generated by choosing $i \sigma_2$ from both $SU(2)$ of the original $SU(2)\times SU(2)$ gauge symmetry, and thus the gauge symmetry after the orbifolding is given by $U(1)\times U(1)\times \mathbb{Z}_2$. This $\mathbb{Z}_2$ symmetry acts as
\begin{align}
y_3\leftrightarrow -y_3, \quad z_1\leftrightarrow -z_1, \quad \psi_3 \leftrightarrow \psi_3, \quad \psi_1 \leftrightarrow -\psi_1, \quad a_{\mu}\leftrightarrow -a_{\mu}, \quad \hat{a}_{\mu}\leftrightarrow -\hat{a}_{\mu}.
\label{gaugedZ2}
\end{align}

In addition to this gauged $\mathbb{Z}_2$-symmetry, we have two more global $\mathbb{Z}_2$-symmetries.
The first is the parity invariance
\beq
A_{\mu}\leftrightarrow \hat{A}_{\mu}, \quad Y_D\leftrightarrow Y_D^{\dagger},
\quad Z_A\leftrightarrow Z_A^{\dagger}, \quad \Psi_{D+} \leftrightarrow \Gamma^1 \Psi^{\dagger}_{D+},\quad  \Psi_{A-} \leftrightarrow \Gamma^1 \Psi^{\dagger}_{A-},
\eeq
which is essentially the same as the un-orbifolded case \cite{Schwarz,vR}.

We also have another discrete $\mathbb{Z}_2$-symmetry, which does not exist in un-orbifolded theory: 
\beq
y^s\leftrightarrow z^i,\quad \hat{A}_{\mu}\leftrightarrow -\hat{A}_{\mu}.\label{eq:Oduality}
\eeq
We will comment on the significance of this $\mathbb{Z}_2$-symmetry later.

\section{Moduli space}\label{sec:moduli}
\subsection{Moduli space of our theory}
\begin{figure}[h]
\centering{\includegraphics[width=8cm,clip]{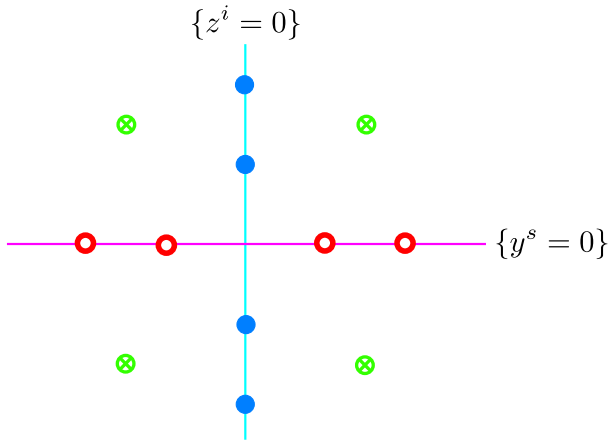}}
\caption{Sketch of solutions (I)-(III). 
M2-branes for phase (I), (II) and (III) are represented by blue, red and green dots, respectively.}
\label{M2position}
\end{figure}

We will now study the moduli space of our model.
In the previous section, we computed the potential 
$V(Z_A,Y_D)$ in (\ref{potential}).
The solutions to $V(Z_A,Y_D)=0$ are classified into the three phases.
\begin{eqnarray}
&({\rm I})& z_1^i=0,\quad z_2^i=0,\quad (i=1,2,3,4),
\\
&({\rm II})& y_3^s=0,\quad y_4^s=0,\quad (s=5,6,7,8),
\\
&({\rm III})& y_3^sy_4^t=y_4^sy_3^t,\quad z_1^iz_2^j=z_2^iz_1^j.
\end{eqnarray}
The corresponding configurations of M2-branes are shown in Fig.~\ref{M2position}. At generic point of moduli space (phase (III)), we have essentially a single M2-branes together with its three mirror images. When M2-branes lies at the fixed locus of $\mathbb{Z}_2$ (phase (I) and phase (II)), we have two M2-branes confined to fixed locus, together with their mirror images.

\noindent \underline{Phase ({\rm I}): M2 at the fixed locus of the orbifold $\mathbb{Z}_2$} \\
In this case, the solution for $V(Z_A,Y_D)=0$ is
\begin{eqnarray}
Z_A^i=0, \quad & Y_D^s
=\left(
\begin{array}{cc}
y^s & 0\\
0 & \bar{y}^s
\end{array}
\right),
\end{eqnarray}
where $y:=y_4+iy_3$. 
%

To find the moduli space, we have to take into account 
$U(1)\times U(1)\times \mathbb{Z}_2$ gauge symmetry.  

Naively we can use $U(1)\times U(1)$ gauge symmetry
and fix one of the phases of $y^s$'s.
Here we simply assume without justification that
the $U(1)$ gauge symmetry coupled to $y$ is broken to a discrete 
subgroup $\mathbb{Z}_{m}$ where $m$ is a some integer number. 
\footnote{There are subtleties in this 
argument. 
We cannot apply the mechanism in 
\cite{Mfold,LT}
via the dual photon, because both of the 
$U(1)\times U(1)$ gauge fields
$b_{\mu}:=a_{\mu}-\hat{a}_{\mu}$ and $c_{\mu}:=a_{\mu}+\hat{a}_{\mu}$ 
couple to the scalar fields in the action 
(\ref{comp action}) 
and the auxiliary fields cannot be introduced.
Within the framework of the Bagger-Lambert theory, we could not justify
this point explicitly. But we expect such mechanism happens, 
because the matching of the moduli spaces of M-theory and Type IIA
for the each branches should be realized.
}
This $\mathbb{Z}_{m}$
acts on $y^s$ as 
\begin{eqnarray}
y^s\to e^{2\pi n i  /m}y^s.
\label{Z_k}
\end{eqnarray}
We also have the gauged $\mathbb{Z}_2$-symmetry \eqref{gaugedZ2}
\begin{align}
y^s\to \bar{y}^s.
\end{align}

Combining these, we have the dihedral group $D_{m}
=\mathbb{Z}_2 \ltimes \mathbb{Z}_{m}$ and the resulting moduli
space is given by
%
\begin{eqnarray}
{\cal M}^{(I),m}= (\mathbb{R}^4\times \mathbb{R}^4)/D_{m},
\label{moduli(i)}
\end{eqnarray}
and the unbroken gauge symmetry is $U(1)_V$, which is generated by $a_{\mu}+\hat{a}_{\mu}$. In the special case $m=4$, we have
\begin{eqnarray}
{\cal M}^{(I), \,m=4}
= \frac{(\mathbb{R}^4/\mathbb{Z}_2)  \times (\mathbb{R}^4/\mathbb{Z}_2)}{\mathbb{Z}_2}.
\label{moduli(i)-1}
\end{eqnarray}
\\

\noindent\underline{Phase (II): M2 at the other fixed locus}\\
In this case, the solution for $V(Z_A,Y_D)=0$ is
\begin{eqnarray}
Y_D^s=0, \quad & Z_A^i
=\left(
\begin{array}{cc}
0 & z^i\\
-\bar{z}^i & 0
\end{array}
\right),
\end{eqnarray}
where $z:=z_2+iz_1$.
%
%
Due to the presence of $\mathbb{Z}_2$-symmetry \eqref{eq:Oduality}, 
 we find that the moduli space for phase (II) 
is isomorphic to that of phase (I):
\begin{eqnarray}
{\cal M}^{(I),m}\simeq 
{\cal M}^{(II),m}.
\end{eqnarray}
The unbroken gauge symmetry is $U(1)_A$,\footnote{As a special point of phase (I) and (II), i.e. when $y^s$ and $z^i$ are all equal to zero, the unbroken gauge group is enhanced to $U(1)_V\times U(1)_A$.} which is generated by $a_{\mu}-\hat{a}_{\mu}$.\\

\noindent \underline{Phase (III): Generic point in moduli space}:\\
In this case, the general solution for the $V(Z_A,Y_D)=0$ is 
\begin{eqnarray}
&&Z_A^i=\left(
\begin{array}{cc}
0 & z_0^ie^{i\phi}  \\
-z_0^ie^{-i\phi} &0
\end{array}
\right),
\quad 
Y_D^s=\left(
\begin{array}{cc}
y_0^se^{i\theta} & 0 \\
0 & y_0^se^{-i\theta}
\end{array}
\right),
\quad 
\end{eqnarray}
where $z_0^i$ and $y_0^s$ are real. 
There are discrete $\mathbb{Z}_2\times\mathbb{Z}_2$ symmetries
\begin{eqnarray}
(z^i,\phi)\to (-z^i,\phi+\pi),\quad
(y^s,\theta)\to (-y^s,\theta+\pi).
\end{eqnarray}
By using $U(1)\times U(1)$ symmetry we can fix the phases 
$\theta$ and $\phi$ to be zero. 

In this vacuum there are no residual symmetries in gauge fields
in contrast to phases (I) and (II).
Actually the action for scalars and gauge fields are given by
\begin{equation}
\begin{split}
{\cal L}_{g}=&-\frac{1}{4}
|\partial_{\mu}y^s+i(a_{\mu}-\hat{a}_{\mu})y^s|^2
-\frac{1}{4}
|\partial_{\mu}z^i+i(a_{\mu}+\hat{a}_{\mu})z^i|^2 \\
&+\frac{k^{\prime}}{2\pi}\epsilon^{\mu\nu\rho}
(a_{\mu}\partial_{\nu}a_{\rho}-\hat{a}_{\mu}\partial_{\nu}\hat{a}_{\rho}).
\end{split}
\end{equation}
For the generic point in the moduli space $y^s\ne 0$, $z^i\ne 0$,
the minimum of this action is realized for $a_{\mu}=\hat{a}_{\mu}=0$.
Then the moduli space for this case consists only of scalar fields.

As a result we find the moduli space ${\cal M}^{(III)}$
for phase (III) is
\begin{eqnarray}
{\cal M}^{(III)}=(\mathbb{R}^4/\mathbb{Z}_2)\times (\mathbb{R}^4/\mathbb{Z}_2).
\label{PhaseIII}
\end{eqnarray}
This result is independent of $k^{\prime}$.

\subsection{Comparison with Type IIA moduli space}

We are now in a position to compare the moduli space of our theory obtained so far to that of D2-branes in the strong coupling limit of Type IIA string theory. If our theory really describes theories on membranes, then 
these two moduli spaces should match. This serves as a good consistency
check of Bagger-Lambert theory and our $\mathbb{Z}_2$-orbifolding
procedure. At first sight the analyses in M-theory and Type IIA look
similar, but at closer inspections of field contents in two theories are
largely different and the match is far from trivial.

The discussion of $\mathbb{Z}_2$-orbifolding of $O(4)$
gauge theory\footnote{
In \cite{LT,Mfold} 
the Type IIA string theory configuration corresponding to 
the un-orbifolded theory with $k=1$ 
is discussed.
Via Higgsing, they found that the Type IIA moduli space for $k=1$
describes the configuration of 
one O2$^-$-plane and two D2-branes (together with their mirror
images).
The resulting worldvolume theory is $SO(4)$ gauge theory rather than 
$O(4)$.
Actually the $O(4)$ gauge theory is 
found naturally in \cite{ABJM,K-theory}.} is analogous to the discussion above of the M-theory case. Using $4\times 4$ matrix representations, take the regular representation $\gamma$ to be
\begin{align}
\gamma=\left(
\begin{array}{cc|cc}
1 & 0 &0 &0 \\
0 & 1 &0 &0 \\
\hline
0 & 0 &-1 &0 \\
0 & 0 &0 &-1 
\end{array}
\right),
\end{align}
and consider $\mathbb{Z}_2$-action as in \eqref{eq:Z2}:
\begin{equation}
 Z^i \to -\gamma Z^i \gamma, ~~ Y^{s'} \to \gamma Y^{s'} \gamma, ~~
\Psi \to \Gamma^{1234} \gamma \Psi \gamma,~~
A_{\mu} \to \gamma A_{\mu}\gamma, 
\label{eq:Z2-2}
\end{equation}
where the seven scalars are decomposed into four scalars $Z^i\, (i=1,2,3,4)$ and $Y^{s'}\, (s'=5,6,7)$. Here we are taking the M-theory direction to be the 8-direction. By this $\mathbb{Z}_2$-action, the remaining fields are $Y^{s'}_D, Z^i_A, \Psi_{D+}, \Psi_{A-}$ and $A_{D\mu}$, where suffixes $D$ (and $A$) represents $2\times 2$ block diagonal (block off-digonal) components. 
For example, gauge field $A_{D\mu}$ after the $\mathbb{Z}_2$-truncation is represented by
\begin{equation}
A_{D\mu}=\left(
\begin{array}{cc|cc}
0&a^A_{\mu} &0 &0 \\
-a^A_{\mu} &0 &0 &0 \\
\hline
0&0 &0 &a^V_{\mu} \\
0&0 &-a^V_{\mu} &0 \\
\end{array}
\right),
\end{equation}
where (up to irrelevant coefficients) in our previous notation in the M-theory,  we have written $a^V_{\mu}=a_{\mu}+\hat{a}_{\mu}$ and $a^A_{\mu}=a_{\mu}-\hat{a}_{\mu}$.
After orbifolding, the gauge symmetry is given by $SO(2)\times SO(2)\simeq U(1)\times U(1)$, plus discrete gauge symmetries which we will comment on in a moment.

The moduli space of this theory again consists of three branches:
\begin{align}
&({\rm i}): Y^{s'}\ne 0, Z^i=0, \\
&({\rm ii}): Y^{s'}= 0, Z^i\ne 0, \\
&({\rm iii}): Y^{s'}\ne , Z^i\ne 0.
\end{align}
The corresponding configurations of D2-branes are almost the same as in M-theory case, namely as in Figure \ref{M2position}. The only difference is that we have only three $Y^{s'}$ directions, not four. We now analyze each phase in detail. \\

\noindent \underline{Phase (i): D2 at the fixed locus of the orbifold $\mathbb{Z}_2$}

In this phase, only the $Y^{s'}$'s take non-zero value:
\begin{align}
Y^{s'}=\left(\begin{array}{cc|cc}
0 & \alpha^{s'} & 0 & 0\\
-\alpha^{s'} & 0 & 0 & 0\\
\hline
0 & 0 & 0& \beta^{s'} \\
0& 0 & -\beta^{s'} & 0
\end{array}\right),\quad
Z^i=0,
\end{align}
where $\alpha^{s'}$ and $\beta^{s'}$ are arbitrary real numbers. 

At this phase, the gauge symmetry $U(1)_V\times U(1)_A$ is completely preserved. This means in addition to scalars $\alpha^{s'}$ and $\beta^{s'}$, we have two periodic parameters $\sigma_V$ and $\sigma_A$ obtained by dualizing two gauge fields $a^A_{\mu}$ and $a^V_{\mu}$. Thus we have $\mathbb{R}^3\times \mathbb{R}^3\times \mathbb{S}^1\times \mathbb{S}^1$, parametrized by $\alpha^{s'},\beta^{s'}, \sigma_V$ and $\sigma_A$. However, we still have to take care of discrete symmetries of $O(4)$. Namely, two discrete symmetries in $SO(4)$
%
\begin{align}
\left(
\begin{array}{cccc}
0 & 1& 0&0 \\
1& 0& 0&0 \\
0 &0 & 0&1 \\
0& 0& 1& 0\\
\end{array}\right), \quad
\left(\begin{array}{cccc}
0 & 0& 1&0 \\
0& 0& 0&1 \\
1 &0 & 0&0 \\
0& 1& 0& 0\\
\end{array}\right),
\label{SO4Z2}
\end{align}
gives two $\mathbb{Z}_2$-symmetries
\begin{align}
\alpha^{s'}\to -\alpha^{s'}, \quad \beta^{s'}\to -\beta^{s'} \quad a^A_{\mu} \to -a^A_{\mu}, \quad \sigma_A \to -\sigma_A, \quad  a^V_{\mu} \to -a^V_{\mu}, \quad  \sigma_V \to -\sigma_V,
\end{align}
and
\begin{align}
\alpha^{s'}\to \beta^{s'}, \quad a^V_{\mu} \to a^A_{\mu}, \quad  \sigma_V \to \sigma_A,
\end{align}
while keeping other fields fixed. Further, discrete symmetry in $O(4)$
\begin{align}
\left(\begin{array}{cccc}
-1 & 0& 0&0 \\
0& 1& 0&0 \\
0 &0 & 1&0 \\
0& 0& 0& 1\\
\end{array}\right),
\label{O4Z2}
\end{align}
gives one more $\mathbb{Z}_2$-symmetry
\begin{align}
\alpha^{s'}\to -\alpha^{s'}, \quad a^A_{\mu} \to -a^A_{\mu}, \quad \sigma_A \to -\sigma_A.
\end{align}
Combining all these three discrete $\mathbb{Z}_2$, the moduli space is given by
\begin{align}
\mathcal{M}^{(i)}=\frac{((\mathbb{R}^3\times \mathbb{S}^1)/ \mathbb{Z}_2)\times ((\mathbb{R}^3\times \mathbb{S}^1)/ \mathbb{Z}_2) }{ \mathbb{Z}_2}
\end{align}
When the coupling goes to infinite, $\mathbb{S}^1$ decompactify\footnote{According to the interpretation of \cite{LT,Mfold}, expectation values of $X_I$'s represent the location of M2-brane in the uncompactified M-theory, not the compactification radius as in \cite{M2toD2}.}
 and we have the correct moduli space
 $((\mathbb{R}^4/\mathbb{Z}_2)\times
 (\mathbb{R}^4/\mathbb{Z}_2))/\mathbb{Z}_2$, as expected\footnote{If we
 use $SO(4)$ gauge group rather than $O(4)$, one $\mathbb{Z}_2$ factor
 is unnecessary and the moduli space becomes
 $(\mathbb{R}^4/\mathbb{Z}_2)^2$. 
This is consistent with the phase (I) moduli space of M-theory with
 $m=4$. Although the breaking of $U(1)_A$ symmetry in
M-theory could not be explained in the context of the orbifolding 
for the Bagger-Lambert theory, this fact will support our assumption. The same discussion applies to phase (II) and (ii) as well. In phase (iii), however, if we use $SO(4)$ gauge group we have branches (as we will see in \eqref{pm}), and the moduli space seemingly does not match with that of phase (III).}:
\begin{align}
\mathcal{M}^{(i)} \to \mathcal{M}^{(I),m=4}, \quad \mbox{as} \quad g_{YM}\to \infty.
\end{align}
\\

\noindent \underline{Phase (ii): D2 on the orientifold}

In M-theory, moduli of Phase (I) and that of Phase (II) are automatically isomorphic, due to the presence of discrete $\mathbb{Z}_2$-symmetry \eqref{eq:Oduality}. It is non-trivial, however, to verify the corresponding fact for Type IIA, because orbifold and orientifold are different in Type IIA.

In phase (ii), the scalars are given by
\begin{align}
Y^{s'}= 0,
\quad
Z^i=\left(\begin{array}{cc|cc}
0 & 0 & \gamma^i & 0\\
0 & 0 & 0 & \delta^i\\
\hline
-\gamma^i & 0 & 0& 0\\
0& -\delta^i & 0 & 0
\end{array}\right),
\end{align}
where $\gamma^i$ and $\delta^i$ are real numbers. 
The form of $Z^i$'s are chosen so that $Z^i$'s mutually commute, thereby minimizing the potential. On this phase, the gauge symmetry is completely broken\footnote{Gauge symmetry $U(1)_V$ (resp. $U(1)_A$) is restored, however, when $\gamma^i=\delta^i$ (resp. $\gamma^i=-\delta^i$).} and we have no scalars coming from the gauge field. By taking care of discrete gauge transformations \eqref{SO4Z2} and \eqref{O4Z2}, 
we have three $\mathbb{Z}_2$-identifications 
(1) $\gamma^i\leftrightarrow -\gamma^i,  \delta^i \leftrightarrow \delta^i$, 
(2) $\gamma^i\leftrightarrow \gamma^i,  \delta^i \leftrightarrow -\delta^i$,
(3) $\gamma^i\leftrightarrow \delta^i$, 
and thus we have the moduli space
\begin{align}
\mathcal{M}^{(ii)}=\frac{(\mathbb{R}^4/\mathbb{Z}_2) \times (\mathbb{R}^4/\mathbb{Z}_2)}{\mathbb{Z}_2}=
\mathcal{M}^{(II),m=4}. 
\end{align}
In this case, the moduli space coincides with that of M-theory even before taking the strong gauge coupling limit.\\

\noindent \underline{Phase (iii): D2 at the generic point of the moduli space}

In this phase, both $Y^{s'}$'s and $Z^a$'s take non-zero value:
\begin{align}
Y^{s'}=\left(\begin{array}{cc|cc}
0 & \alpha^{s'} & 0 & 0\\
-\alpha^{s'} & 0 & 0 & 0\\
\hline
0 & 0 & 0& \beta^{s'} \\
0& 0 & -\beta^{s'} & 0
\end{array}\right),\quad
Z^i=\left(\begin{array}{cc|cc}
0 & 0 & \gamma^i & 0\\
0 & 0 & 0 & \delta^i\\
\hline
-\gamma^i & 0 & 0& 0\\
0& -\delta^i & 0 & 0
\end{array}\right),
\end{align}
In order to minimize the potential, these matrices should commute, giving us the condition
\begin{align}
\alpha^{s'} \gamma^i=\beta^{s'} \delta^i,\quad
\alpha^{s'} \delta^i=\beta^{s'} \gamma^i,\quad
\end{align}
which given us
\begin{align}
\alpha^{s'}=\pm \beta^{s'},\quad 
\gamma^{i}=\pm \delta^{i},
\label{pm}
\end{align}
where we should take the same sign for two equations in \eqref{pm}. In this phase, the unbroken gauge symmetry is given by $U(1)_V$ (resp. $U(1)_A$) when we take the plus (resp. minus) sign in \eqref{pm}. This contributes one extra scalar $\sigma_V$ (resp. $\sigma_A$) to the moduli space.

Again by taking care of discrete gauge symmetries, the two choices of $\pm$ in \eqref{pm} are identified by \eqref{O4Z2}, and we have in addition two discrete gauge symmetries
\begin{align}
\alpha^{s'}\to -\alpha^{s'},\quad \sigma_V\to -\sigma_V,
\end{align}
and
\begin{align}
\delta^{i}\to -\delta^{i}.
\end{align}
We thus have
\begin{align}
\mathcal{M}^{(iii)}=
\frac{\mathbb{R}^4}{\mathbb{Z}_2} \times \frac{\mathbb{R}^3\times \mathbb{S}^1}{\mathbb{Z}_2}.
\end{align}

When we go to the strong gauge coupling limit, $\mathbb{S}^1$ again decompactify and we thus have the moduli space $(\mathbb{R}^4/\mathbb{Z}_2 )\times (\mathbb{R}^4/\mathbb{Z}_2 )$, which is consistent with the M-theory analysis in \eqref{PhaseIII}.\\


\section{Conclusions and Discussions}\label{sec:conclusion}

In this paper, we have proposed a new Lagrangian describing
$\mathcal{N}=4$ superconformal field theory in three dimensions. This
Lagrangian is likely to describe interacting field theory on the worldvolume of a M2-brane placed on an orbifold $\mathbb{R}^8/(\mathbb{Z}_2\times \mathbb{Z}_2$), and is obtained as a $\mathbb{Z}_2$-orbifold of Bagger-Lambert theory in the $SU(2)\times SU(2)$ bifundamental representations. 

We also analyzed the moduli space of our theory and found three
branches.  
In the analysis of the Phase (I) and (II), we assumed some mechanism to
make one of $U(1)$ gauge symmetry be broken to the discrete subgroup 
$\mathbb{Z}_{m}$. Within the framework of the Bagger-Lambert
theory, we could not justify this mechanism explicitly.
But under this assumption, the matching of the moduli 
spaces of M-theory and Type IIA theory for each branches can be found 
especially for $m=4$ in highly non-trivial way.
In this discussion, the moduli space for the Type IIA theory is 
given by the $\mathbb{Z}_2$-orbifold of $O(4)$ gauge theory, 
rather than $SO(4)$ as in
\cite{Mfold,LT}.
Conversely speaking, the M2-brane theory on the $\mathbb{Z}_2$-orbifold
should be defined as the strong coupling limit of the Type IIA brane
configuration on $\mathbb{Z}_2$-orbifold, then the matching of 
each branches of moduli space supports our assumption and analysis in M-theory.

The interesting feature of our Lagrangian is the existence of $\mathbb{Z}_2$-symmetry \eqref{eq:Oduality}, which exchanges two $\mathbb{Z}_2$-actions. In M-theory viewpoint this is natural and simplify exchanges two $\mathbb{Z}_2$-actions, but in Type IIA language this exchanges orbifold with orientifold, which is highly non-trivial. In our discussion, we have deleted 8-direction (i.e. one of the $Y^s$-directions) to obtain $\mathbb{Z}_2$-orbifold of D2-O2$^-$ system. If we instead reduce along $Z^i$-directions, then we should have D6-D2-O2$^-$ system without $\mathbb{Z}_2$-orbifold.  
Now the symmetry \eqref{eq:Oduality} implies a new duality between $\mathbb{Z}_2$-orbifold of O2$^-$ and D6-O2$^-$. We call this new non-perturbative duality ``O-duality''.\footnote{O stands for orientifold and orbifold, and also for gauge groups $O(N)$.}  \footnote{The existence of duality is not limited to BLG theory and exists also in the orbifold of $U(2)\times U(2)$ ABJM theory \cite{ABJM}, as discussed in section 4.2 of \cite{TY}. In the notation of the paper, the $\mathbb{Z}_2$-symmetry exchanges $Z^1,W^1$ and $Z^2,W^2$.} The existence of orientifold is crucial for the existence of this duality. As a possible check of this proposal, our moduli space in phase (I) should match with the instanton moduli space of $SU(2)$-instanton placed at an $\mathbb{Z}_2$-orbifold, and it would be interesting to explicitly verify this.

Finally, in this paper we have concentrated on a single example of $\mathbb{Z}_2$ acting on $\mathbb{R}^4$. We can consider more examples by considering 
$\mathbb{Z}_2$ acting on $\mathbb{R}^2$, $\mathbb{R}^6$ and $\mathbb{R}^8$, for example, and it would be interesting to study them.

\section*{Acknowlegments}
We would like to thank Futoshi Yagi for discussions and collaborations in early stages of this project. M.~Y.~would like to thank Yukawa Institute for Theoretical Physics for hospitality during this work. H.~F.~is supported by JSPS Grant-in-Aid for Creative Scientific Research, No. 19GS0219. S.~T.~is partly supported by
the Japan Ministry of Education, Culture, Sports, Science and
Technology. M.~Y.~is supported in part by JSPS fellowships for Young Scientists.

\appendix
\section{Notations of $\Gamma$-matrices}\label{sec:gamma}

In this appendix, we explain the origin of the $\Gamma^{1234}$ factor in \eqref{eq:Z2}. 

For our purpose, it is convenient to use the following explicit representations of the 11-dimensional $\Gamma$-matrices:

\begin{equation}
\begin{split}
\Gamma^1&=1\times \tau_3\times \ep \times \ep \times \tau_3, ~~
\Gamma^2=\tau_1 \times \ep \times 1\times \ep \times \tau_3, \\
\Gamma^3&=\tau_3\times \ep \times 1 \times \ep \times \tau_3, ~~
\Gamma^4=\ep \times 1 \times \tau_1 \times \ep \times \tau_3, \\
\Gamma^5&=1 \times \tau_1\times \ep \times \ep \times \tau_3, ~~
\Gamma^6=\ep \times \ep\times \ep \times \ep \times \tau_3, \\
\Gamma^7&=\ep\times 1\times \tau_3 \times \ep \times \tau_3, ~~
\Gamma^8=1\times 1\times 1\times \tau_1\times \tau_3, \\
\Gamma^9&=1\times 1\times 1\times 1\times \tau_1, ~~
\Gamma^0=1\times 1\times 1\times 1\times \ep,\\
\Gamma^{10}&=\Gamma^0\Gamma^1\ldots \Gamma^9=1 \times 1 \times 1\times \tau_3 \times \tau_3
\end{split}
\label{eq:gammarepr}
\end{equation}

We want to study the effect of reflections ($Z^i\to -Z^i, Y^s\to Y^s$) on the fermion $\Psi$. This $\mathbb{Z}_2$-action is equivalent to $\pi$ rotations in 12-planes and 34-planes. In the representation of \eqref{eq:gammarepr}, generators of rotations in 12- and 34- planes are given by
\begin{align}
\Sigma_{12}=\frac{-i}{4} [\Gamma^1,\Gamma^2], ~~
\Sigma_{34}=\frac{-i}{4} [\Gamma^3,\Gamma^4], 
\end{align}
with 
\begin{align}
\frac{1}{2} [ \Gamma^1,\Gamma^2 ]=\epsilon\times 1\times 1\times 1\times 1, ~~
\frac{1}{2} [ \Gamma^3,\Gamma^4 ]= 1\times 1\times \epsilon\times 1\times 1.
\end{align}

By using the identity
\begin{align}
\exp\left(\frac{\pi}{2}\ep\right) =\cos\left(\frac{\pi}{2}\right)\, \mathbf{1}+\sin \left(\frac{\pi}{2}\right) \, \epsilon =\ep,
\end{align}
we obtain
\begin{align}
\exp\left(i \pi (\Sigma_{12}+\Sigma_{34})\right)=\ep\ti 1\ti \ep \ti 1\ti 1 =\Gamma^1 \Gamma^2 \Gamma^3 \Gamma^4=\Gamma^{1234},
\end{align}
and we find $\Gamma^{1234}$ factor in \eqref{eq:Z2}, as expected. 
Note that the final result is independent of specific representations of $\Gamma$-matrices we used above.

\end{document}